\documentclass[useAMS,usenatbib]{mn2e}
\usepackage{graphicx}
\usepackage{times}
\usepackage[dvips]{color}

\title{Disentangling the Galaxy at low Galactic latitudes}

\author[M.\,Cignoni, M.\,Tosi, A.\,Bragaglia, J.S.\,Kalirai, D.S.\,Davis]{M.\,
       Cignoni$^{1,2}$\thanks{E-mail:michele.cignoni@unibo.it},
       M.\,Tosi$^{2}$, A.\,Bragaglia$^{2}$,
       J.S.\,Kalirai$^{3}$, D.S.\,Davis$^{4}$\\ $^{1}$Dipartimento di
       Astronomia, Universit\`a degli Studi di Bologna, via Ranzani 1, I-40127
       Bologna, Italy\\ $^{2}$Istituto Nazionale di Astrofisica, Osservatorio
       Astronomico di Bologna, Via Ranzani 1, I-40127 Bologna, Italy\\
       $^{3}$UCO/Lick Observatory, Department of Astronomy and Astrophysics,
       University of California at Santa Cruz, 1156 High Street, Santa Cruz,
       CA 95064\\ $^{4}$Department of Physics and Astronomy, University of
       British Columbia, 6224 Agricultural Road, Vancouver, B.C., Canada.}

\begin{document}
      \date{Received; accepted}
\pagerange{\pageref{firstpage}--\pageref{lastpage}} \pubyear{2008}

\maketitle

\label{firstpage}

\begin{abstract}

{We have used the field stars from the open cluster survey BOCCE, to study
  three low-latitude fields imaged with the Canada-France-Hawaii telescope
  (CFHT), with the aim of better understanding the Galactic structure in those
  directions. Due to the deep and accurate photometry in these fields, they
  provide a powerful discriminant among Galactic structure models. In the
  present paper we discuss if a canonical star count model, expressed in terms
  of thin and thick disc radial scales, thick disc normalization and reddening
  distribution, can explain the observed CMDs. Disc and thick disc are
  described with double exponentials, the spheroid is represented with a De
  Vaucouleurs density law. In order to assess the fit quality of a particular
  set of parameters, the colour distribution and luminosity function of
  synthetic photometry is compared to that of target stars selected from the
  blue sequence of the observed colour-magnitude diagrams.  Through a
  Kolmogorov-Smirnov test we find that the classical decomposition
  halo-thin/thick disc is sufficient to reproduce the observations---no
  additional population is strictly necessary. In terms of solutions common to
  all three fields, we have found a thick disc scale length that is equal to
  (or slightly longer than) the thin disc scale.}  {}
\end{abstract}


\section{Introduction}
In order to reconstruct a coherent picture of our Galaxy, star count models
typically exploit the colour-magnitude diagrams (CMD) from several lines of
sight. The primary goal is to constrain the structure and relative strength of
the various Galactic components. In addition, other quantities such as the star
formation rate (SFR), the initial mass function (IMF), the chemical
composition, and the reddening laws are also tested.

Despite the pioneering successes by Bahcall in the \textquoteright 80s (see
e.g. \citealt {BS}) and the extraordinary amount of precise data available
today, many aspects of the Galactic structure remain ambiguous. The number of
Galactic components (halo, bulge, disc, thick disc, etc..), their chemical
composition, and their origin are widely debated. Recent large-scale surveys
(e.g. SDSS, 2MASS, QUEST) have detected the presence of substructures in the
outer halo, which are taken to be the remnants of disrupted galaxies. For the
disc structures, while there is consensus that most of the thin disc
population has a dissipative history, the thick disc origin remains
contentious. There are (at least) three main processes which are now proposed
to be responsible for thick disc formation: 1) \emph{External origin}---the
stars are accreted from outside, during merging with satellite galaxies, 2)
\emph{Induced event}---the thin disc has been puffed up during close
encounters with satellite galaxies, 3) \emph{Evolutionary event}---the thick
disc settled during the collapse of the proto-galactic cloud, before the thin
disc formation.

Given this uncertain scenario, it is intriguing to learn about more or less
pronounced sequences (see e.g. \citealt{conn}) crossing the CMDs at low
Galactic latitudes. However, fitting these features in a self-consistent
scenario is rather challenging. For instance, an incorrect metallicity and a
complex reddening distribution can both conspire to bias results. Moreover, in
order to detect a possible stellar over-density, it is essential to have at
least a rough idea of the underlying Galactic structure. In other words, one
must know how the ``average'' Galactic CMD should look, especially close to
the Galactic plane. A way out could be to observe symmetrical directions
relative to the plane (see e.g. \citealt{conn}): invoking a north-south symmetry, the observed CMDs
should indicate the presence of a bona fide over-density. However, this option
is fraught with uncertainties as well: is the Galaxy symmetrical? Is the
Galaxy (stellar disc) warped (see e.g. \citealt{lc})?  Can asymmetrical reddening
distributions or stellar chemical gradients mimic asymmetrical star counts?

This paper discusses the capability of a Galactic synthesis model to interpret
the star counts at low Galactic latitudes. Typically, star count models create
a main sequence template, and attempting to recover the underlying
distribution. We make use of an alternate scheme in which we try to translate
the current knowledge of the Galactic populations (thin disc, thick disc and
halo) into synthetic CMDs, and see if they are compatible (and at which
degree) with the observed CMDs. We try to answer the following question: do
the many uncertainties on the Milky Way structure allow to explain the
observed CMDs without invoking anomalies? Our method does not produce unique
scenarios, and furthermore, we argue that one cannot generally infer
unique results.

Taking advantage of the deep and wide-field photometry acquired with the CFH
telescope, whose original targets were open clusters close to the Galactic
plane (\citealt{kali}\,a,b,c, 2007), we are sensitive to disc structures for 
several kpc before being dominated by the halo. These low-latitude regions are 
often avoided by star count analyses for their high obscuration. Hence, the
published results suffer from a bias: most of the investigations are devoted
to the study of the disc scale heights and the halo structure, information
available at intermediate to high Galactic latitudes, whilst the disc scale
lengths are often neglected.
 
The results we find in literature are extremely variable, ranging from 2 kpc
to 5 kpc for the thick disc scale length. Some of these studies provide
evidence for a thick disc/thin disc decomposition with similar scale lengths,
while others do not. For instance, \cite{robin} and others find 2.5 kpc for
the thin disc and 2.8 kpc for the thick disc, \cite{ojha} finds a thin disc
scale length of 2.8 kpc and a thick disc of 3.7 kpc, \cite{larsen} find a
thick disc scale length larger than 4 kpc. From edge-on disc galaxies,
\cite{yoac} find support for thick discs larger than the embedded thin discs,
and \cite{park} argue that the thick disc is not axisymmetrical.

The main issue is whether the thick disc is an independent structure. Although
chemical investigations indicate a different $\alpha$-elements history for the
thick disc, suggesting it is a separate component, most of these studies must
assume a well defined kinematical signature, neglecting stars with
intermediate kinematics. A marked scale height difference between the two
discs (250 pc versus 1 kpc), a well established result, does not exclude a
heating origin. The radial scales could in principle distinguish among
different formation scenarios: N-body simulations suggest that a heating
mechanism can increase the scale height of a population, but it hardly
produces a longer scale length.

The paper is organized as follows. First we introduce the data in section
2. Section 3 gives an overview of the method. In sections 4 and 5, the star
counts in each direction are described in terms of thin and thick disc, and
the reddening distribution along the line of sight is determined. In
section 6, we assess the implications of our findings.

\section{Data}
The data used in this study are in three low latitude fields, obtained with
the CFHT, corresponding to the location of the open clusters NGC6819
[(\textit{l,\,b})$^\circ$ = $(73.98,\,\,+8.48)^\circ$], NGC7789
[(\textit{l,\,b})$^\circ$ = $(115.5,\,\,-5.38)^\circ$] and NGC2099
[(\textit{l,\,b})$^\circ$ = $(177.63,\,\,+3.09)^\circ$]. For a complete
description of the observations and reductions see
\citealt{kali}\,a,b,c. These data, which were originally obtained to study
cluster white dwarfs, represent very deep windows in the thin and thick
disc. We have selected these three fields out of the twenty currently
available from the BOCCE (Bologna Open Clusters Chemical Evolution) project
\citep{brag}, because they are the deepest, widest and cleanest ones.

To select bona fide field stars, we specifically focus our analysis on the
$(V,B-V)$ region below the clusters' main sequences: this region differs for each
field (as shown in the upper panel of Figures \ref{fig1}, \ref{fig2},
\ref{fig3}), due to the various locations of the clusters, reddening
distribution, etc... To increase the sensitivity to the structural parameters,
each region has been further divided into subregions.

The chosen ``grid'' is set up keeping several factors in mind: including red
stars gives a better counting statistics; a narrow colour range shortens the
mass range of the stellar populations, weakening the constraints on SFR and
the IMF; and focusing only on blue stars preserves the B-magnitude
completeness. The bright and faint magnitude limits are chosen to avoid
cluster stars, while guaranteeing sample completeness.

In the directions { (\textit {l,\,b})$^\circ$ = $(115.5,\,\,-5.38)^\circ$
    ({\textit NGC7789}) and (\textit {l,\,b})$^\circ$ =
    $(73.98,\,\,+8.48)^\circ$ ({\textit NGC6819}) the bulk of thin disc stars
    are close to the cluster, and therefore, share similar CMD
    positions. Although this is a strong limitation for the thin disc
    analysis, these data still represent a unique chance to study the thick
    disc structure. In fact, few, if any, of the sources with magnitude
    fainter than $V=18$ (and suitable colours) are likely to be physically
    associated to the clusters. Again, thanks to the excellent photometry, we
    can exploit the CMDs as faint as $V=22$. According to simulations, at this
    magnitude it is possible to trace the radial scale length of the thick
    disc. The situation is different in the anticentre field
    (\textit{l,\,b})$^\circ$ = $(177.63,\,\,+3.09)^\circ$ ({\textit
    NGC2099}). The proximity to the Galactic plane offers a deep snapshot of
    the outer thin disc, but consequently provides little information about
    the thick disc. Combining the three lines of sight is an effective test
    for our Galactic model.

Finally, the CMD density of these stars reflects the matter distribution along
the line of sight: the luminosity function is more sensitive to the Galactic
structure, while the colour distribution is a major discriminant for
age/metallicity/reddening combinations.

\section{The model}
Both the thin and the thick disc components are shaped as double exponentials,
characterized by vertical and radial scales. Because of the low-latitude, our
lines of sight are less informative about the vertical structure. In our
simulations, we make the simplifying assumption that the thick disc scale
height ($H_{thick}$) is 1 kpc, which is comfortably within the literature
range, while the thin disc scale height ($H_{thin}$) is tested for three
characteristic values, namely 200, 250, 300 pc. Both the thick and thin disc
radial scale lengths are allowed to vary freely. Halo and thick disc local
densities are expressed  as a fraction of the thin disc density. In
particular, the local halo fraction is fixed to be 0.0015 (see e.g. \citealt{sieg}), while
the thick disc value is a free parameter.

The stellar halo is characterised by a De Vaucouleurs density law with a half-light radius of 2.6 kpc.  Although our data are marginally sensitive to the
halo structure, simulations indicate that this component is required to
improve the quality of the fit.

In conclusion, our Galactic model relies on four free parameters, namely the
two scale lengths ($L_{thin}$ and $L_{thick}$), $H_{thin}$, and the local
thick disc normalization. The complete list of model ingredients is given in
Table \ref{tab} (for further details see e.g. \citealt{cigno1} and \citealt{cast1}). For each component, the SFR is assumed constant. The recent
SFR of the thin disc is chosen to reproduce the blue edge of the CMD.

\begin{table*}
\begin{center}
\begin{tabular}{lrrr}
 & \multicolumn{1}{c}{Thin disc} & \multicolumn{1}{c}{Thick disc} &
 \multicolumn{1}{c}{Halo} \\
\hline
\hline
Range of SFR(Gyr)& {\Large x}-6& 10-12 & 12 \\
Z& 0.02& 0.001-0.006& 0.0004\\
H-scale(pc) & 200,250,300 & 1000 & /\,\,\,\,\, \\
Radial scale(pc)& {\Large x}& {\Large x}& 2600 \\
$[{\rho}/{\rho_{Thin}}]_{\odot}$&1&{\Large x}&0.0015\\
\hline
\end{tabular}
\end{center}
\caption{Model parameters. The X-symbol indicates a variable quantity. The SFR
is assumed constant in the indicated range.}
\label{tab}
\end{table*}

The IMF is a power law with a Salpeter exponent. Masses and ages are randomly
extracted from the IMF and the SFR, colours are interpolated using the
Pisa evolutionary library \citep{car}. Once the absolute photometry is created,
the line of sight is populated according to the density profiles and a
reddening correction is introduced. To reduce the Poisson noise, the model
CMDs are built from samples ten times larger than the observed ones. For stars brighter than V=22,
photometric errors do not exceed 0.01 mag and the completeness in V is around
80\% (\citealt{kali}-b,c), thus the simulated CMDs can be directly compared
with the data without blurring or corrections.

 Finally, to decide the match quality of a given model, a Kolmogorov-Smirnov
test was carried out to investigate both the colour distribution in each
subregion and the luminosity function of the whole box. Only models giving a
KS-probability larger than 0.001 for each constraint are selected.

\section{Results about thin disc and thick disc structure}
The lower panels of Figures \ref{fig1}, \ref{fig2}, and \ref{fig3} show the
synthetic diagrams which best reproduce the observed CMD of the top panels.
\begin{figure}
\centering
\includegraphics[width=5cm]{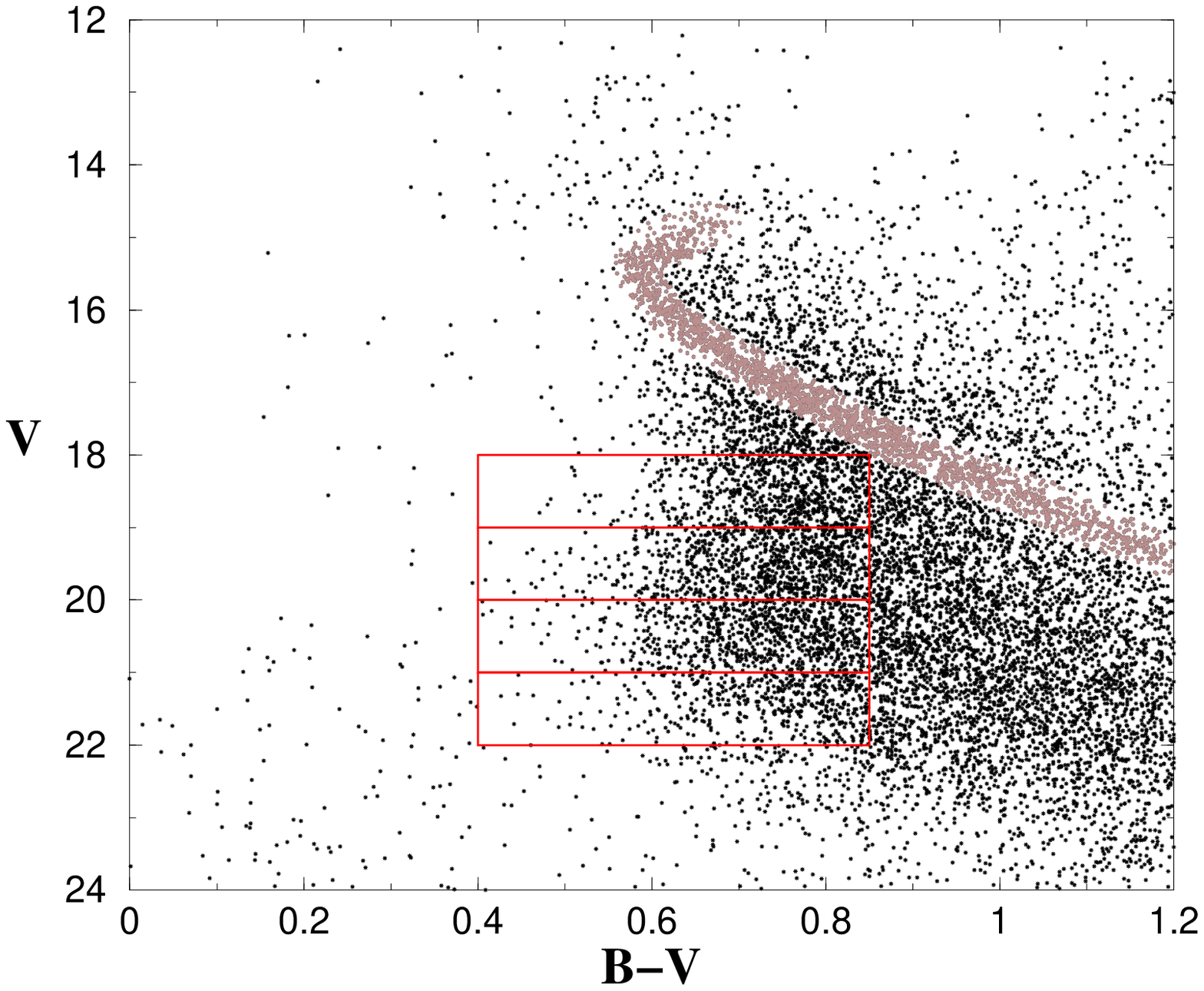}
\includegraphics[width=5cm]{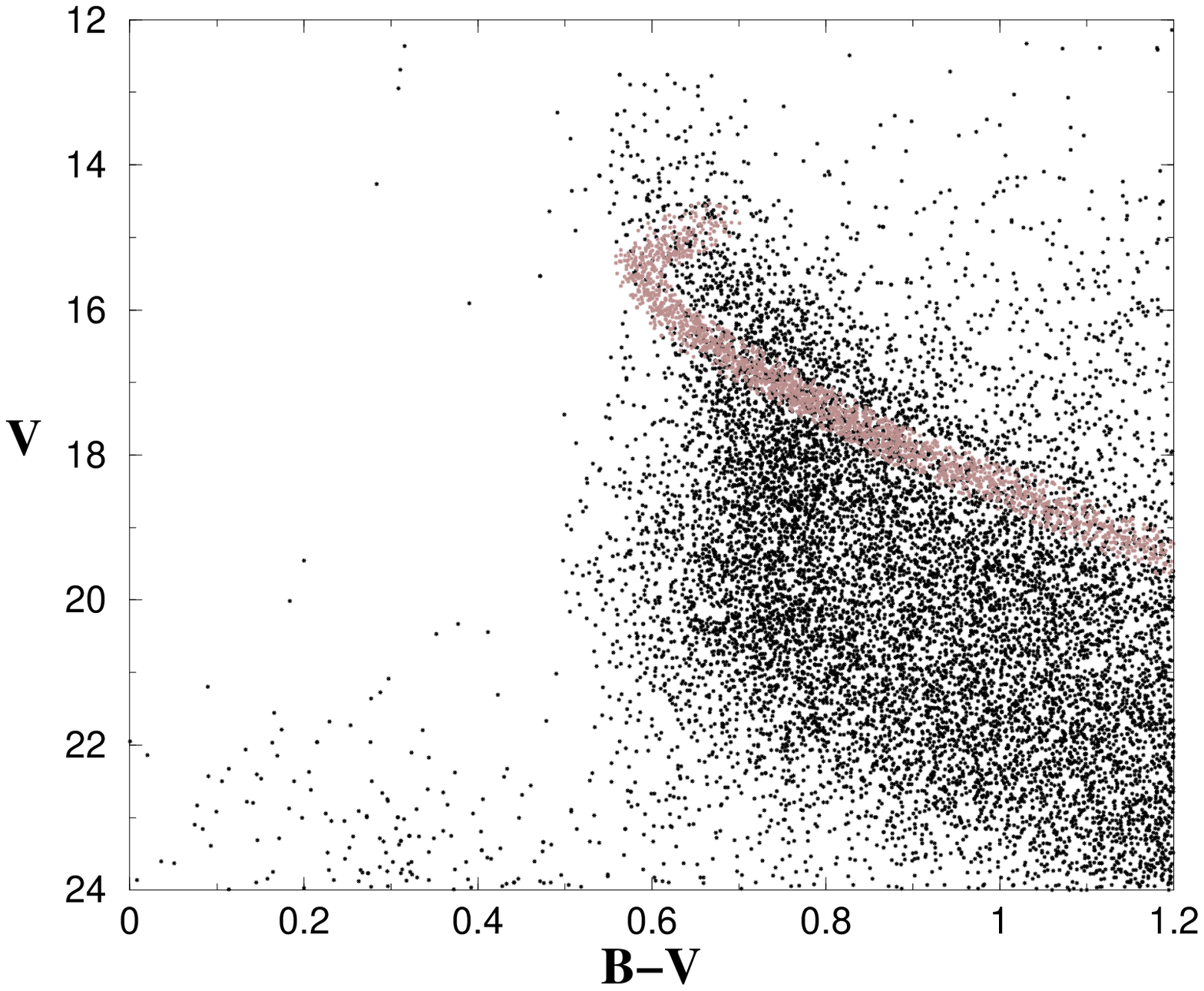}
\caption{Top panel: the observational CMD in the direction of NGC6819
  [(\textit{l,\,b})$^\circ$ = $(73.98,\,\,+8.48)^\circ$] (the cluster main
  sequence is shown in a lighter colour). The field of view is 0.148 square
  degrees. The location of the explored subregions is indicated by the
  boxes. Bottom panel: a synthetic diagram for an acceptable combination of
  parameters. The thin disc scale height is 250 pc, the radial scales for thin
  disc and thick disc are 2500 pc and 3700 pc respectively. The local thick
  disc normalization is 7\%. }
\label{fig1} 
\end{figure}

\begin{figure}
\centering
\includegraphics[width=5cm]{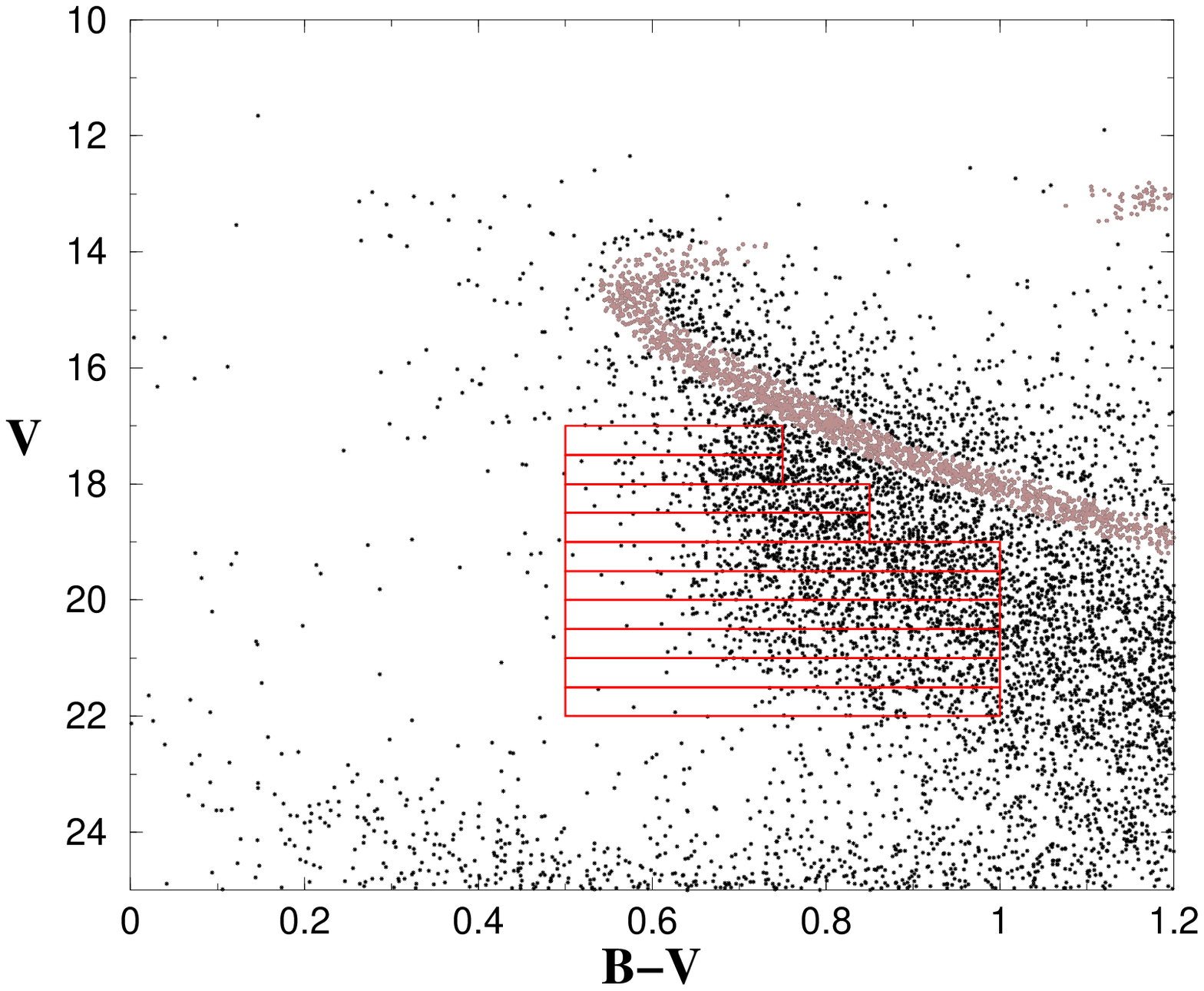}
\includegraphics[width=5cm]{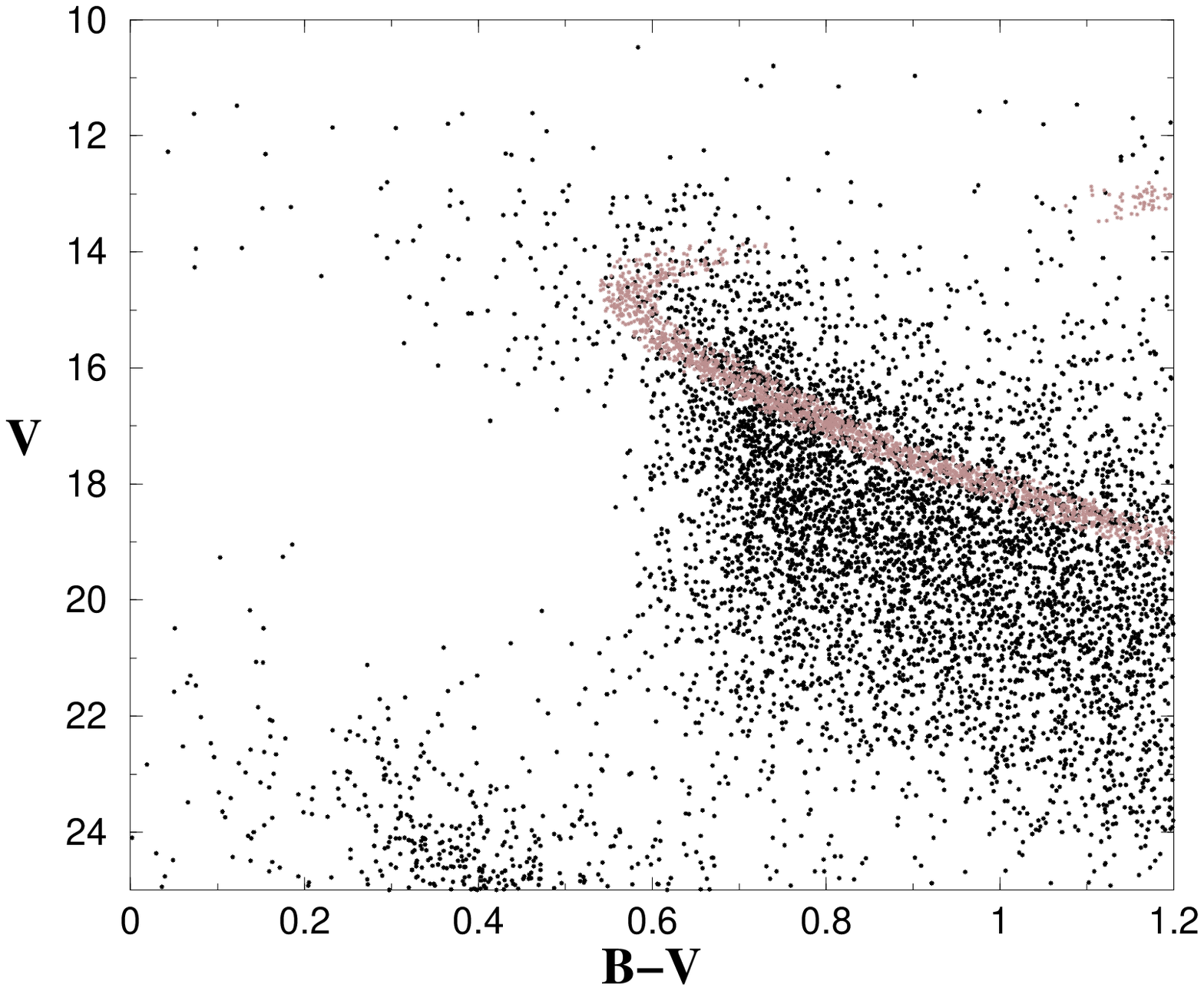}
\caption{As in figure \ref{fig1}, but in the direction of NGC7789
  [(\textit{l,\,b})$^\circ$ = $(115.5,\,\,-5.38)^\circ$]. The field of view is
  0.104 square degrees. Here the thin disc scale height is 250 pc, the radial
  scales for thin disc and thick disc are respectively 2500 pc and 3500
  pc. The local thick disc normalization is 6\%.}
\label{fig2} 
\end{figure}

\begin{figure}
\centering
\includegraphics[width=5cm]{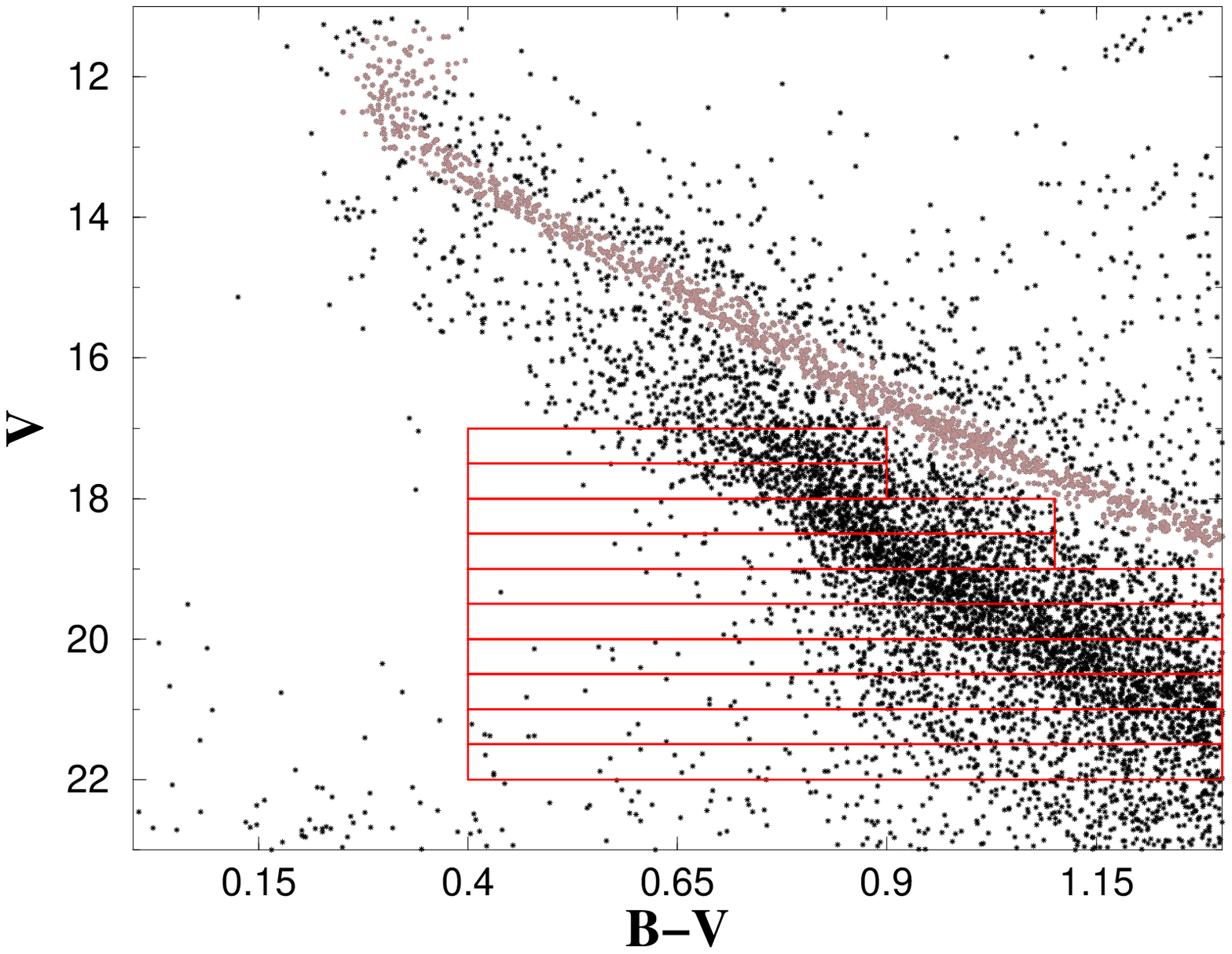}
\includegraphics[width=5cm]{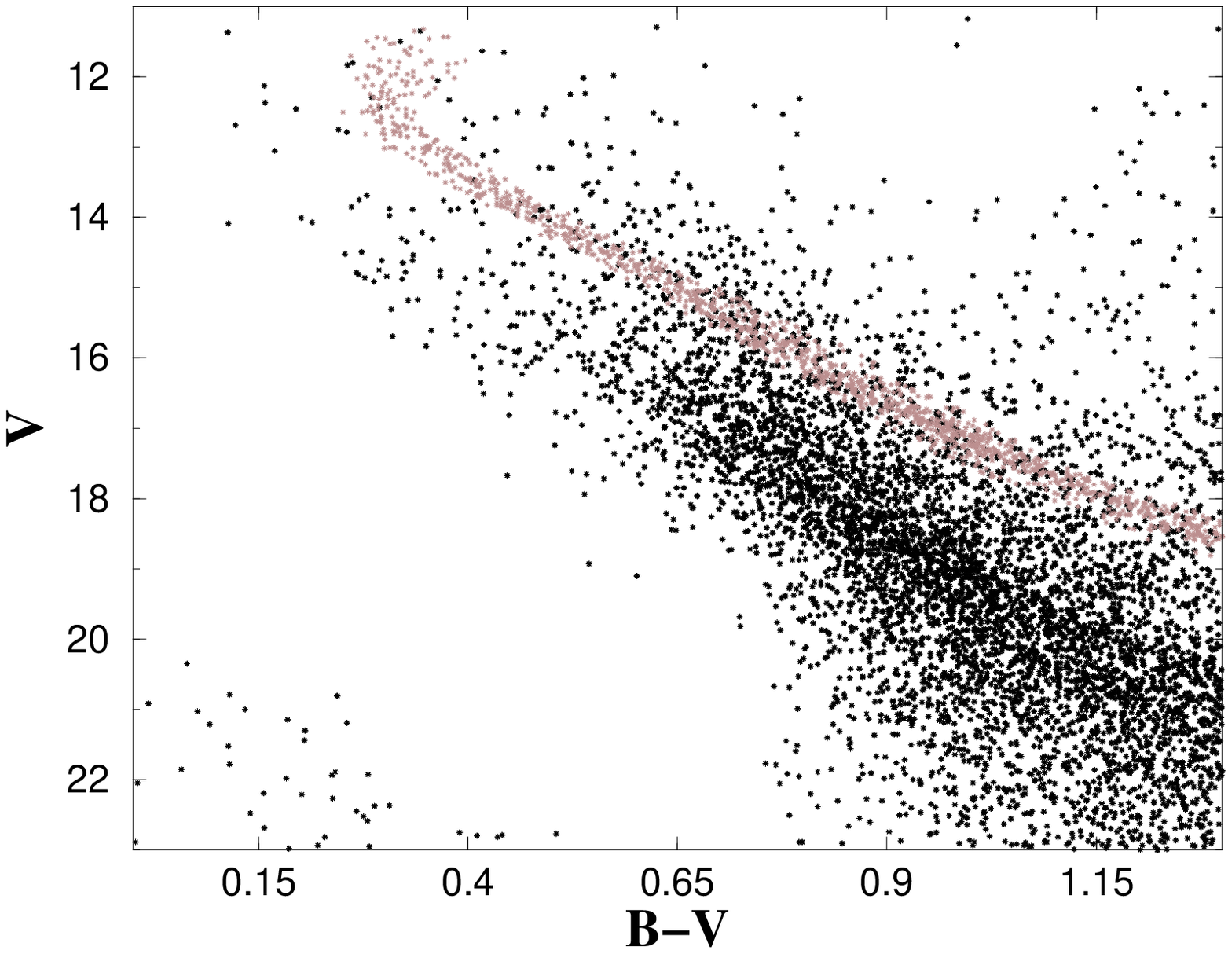}
\caption{As in figure \ref{fig1}, but in the direction of NGC2099
  [(\textit{l,\,b})$^\circ$ = $(177.63,\,\,+3.09)^\circ$]. The field of view
  is 0.155 square degrees. $H_{thin}$ is 250 pc. The radial scales for thin
  disc and thick disc are respectively 2200 pc and 3300 pc. The local thick
  disc normalization is 5\%.}
\label{fig3} 
\end{figure}

Figure \ref{res1} shows the allowed region in the parameter space $L_{thick}$
  versus $L_{thin}$. It seems clear that our fields at (\textit
  {l,\,b})$^\circ$ = $(115.5,\,\,-5.38)^\circ$ ({\textit NGC7789}) and
  (\textit {l,\,b})$^\circ$ = $(73.98,\,\,+8.48)^\circ$ ({\textit NGC6819})
  are poorly suited to constrain the thin disc structure: any thin disc scale
  length between 1000 and 6000 pc looks acceptable. Even allowing for
  different thin disc scale heights does not actually reduce the parameter
  space. In these CMDs, cluster and thin disc are partially overlapped,
  causing the loss of field stars during the already mentioned selection
  process. In conclusion, there are seemingly insufficient thin disc stars in
  our selected regions to determine its properties. In contrast, these directions clearly indicate a preferred range for the thick
  disc scale length. 

The solutions for (\textit{l,\,b})$^\circ$ =
  $(177.63,\,\,+3.09)^\circ$ ({\textit NGC2099}) present the opposite
  situation: in this direction, the $L_{thin}$ values are well constrained,
  while no preferred solution emerges for $L_{thick}$ (which varies
  between 2000 pc and 6000 pc). Evidently, the low latitude of NGC2099
  combined with the intrinsically short vertical scale of the thin disc,
  implies that a significative portion of the thin disc is included in our
  field of view. On the other hand, this is not true for the thick disc, whose
  population density close to the Galactic plain is much lower.

 Remarkably, a region exists in the parameter space which is consistent with
  the combined directions. The three panels of Figure \ref{res1} explore any
  dependence of this region on the thin disc scale height ($H_{thin}$). For
  $H_{thin}=200$ pc the acceptable solutions for $L_{thick}$ do not show any
  correlation with $L_{thin}$ ($L_{thick}$ versus $L_{thin}$ is flat); for
  this scale height, the thin disc has a negligible influence on the explored
  CMD region (the thin disc is brighter than $V\sim 17-18$, i.e., the
  low-magnitude limit). If $H_{thin}$ is increased, the probability of finding
  thin disc stars fainter than $V\sim 17-18$ increases as well. This implies a kind of
  degeneracy between $H_{thin}$ and $L_{thick}$. This effect is particularly
  evident for NGC6819, the highest latitude field: $L_{thick}$ is strongly correlated with $H_{thin}$.

Regardless of the particular thin disc scale height, it is noteworthy that the
  recovered ratio $L_{thick}/L_{thin}$ never falls below one (the straight
  line in Figures \ref{res1} stands for $L_{thick}/L_{thin}=1$). Most of the
  common solutions clump around $H_{thin}=200$ pc, supporting similar values
  for $L_{thick}$ and $L_{thin}$. If $H_{thin}$ is increased, the accepted
  $L_{thick}$ becomes slightly larger than $L_{thin}$.

Figures \ref{res2} show the acceptable pairs of thick disc normalization and
$L_{thick}$. The two parameters influence the luminosity function in a
different fashion. The first parameter is the fraction of thick disc stars
with respect to thin disc stars in the solar neighbourhood: it controls the
ratio between bright (essentially thin disc) and faint (essentially thick
disc) stars. On the other hand, the thick disc scale length is associated both
with the ratio bright/faint \emph{and} with the luminosity function decline in
the faint end (which is always thick disc dominated).

For each field, the solution space of Figures \ref{res2} demonstrates the
strong anti-correlation between $L_{thick}$ and the thick disc local
normalization. This effect is related to the fact that the total number of
stars subtended by an exponential distribution is proportional to the scale;
hence, when the model scale decreases the local normalization must increase,
in order to reproduce the observed starcounts. Common solutions exist only for
$H_{thin}$ shorter than 250 pc. Beyond this value, the space solutions for the
directions to NGC6819 and NGC7789 split into two distinct regions and common
solutions no longer exist.  Summarising:
\begin{enumerate}
\item{ the thin disc scale height is shorter than 250 pc;}
\item {the thick and the thin disc scale lengths are similar;}
\item{the scale lengths are quite short (2250-3000 pc for the thin disc,
  2500-3250 pc for the thick disc);}
\item {a set of parameters ($\rho_{thick}, L_{thick}, L_{thin},H_{thin}$) can
  \emph{simultaneously} satisfy the requirements of the three directions;}
\item {the thick disc normalization is smaller than 10\%.}
\end{enumerate}

\begin{figure}
\centering
\includegraphics[width=7cm]{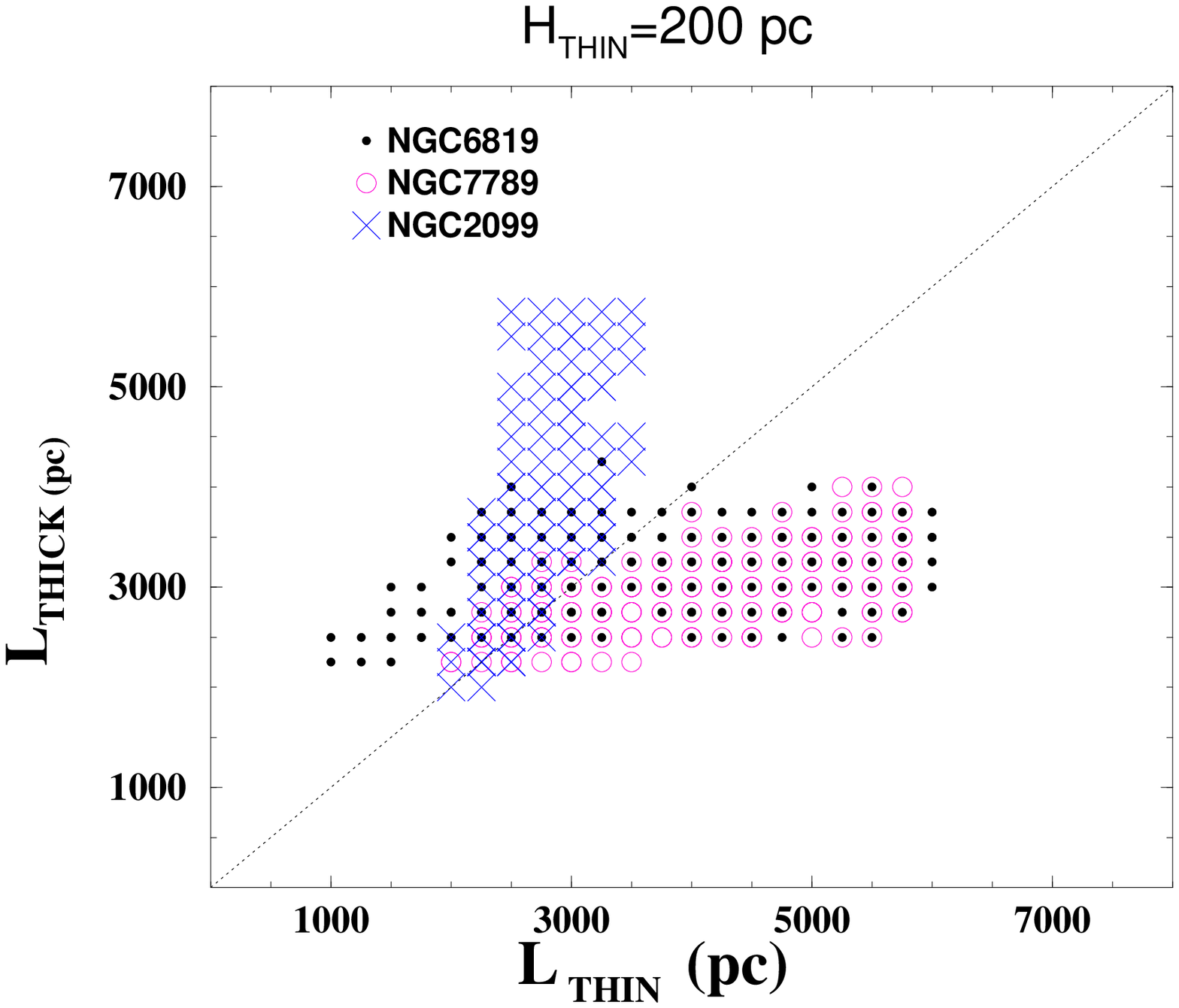}
\includegraphics[width=7cm]{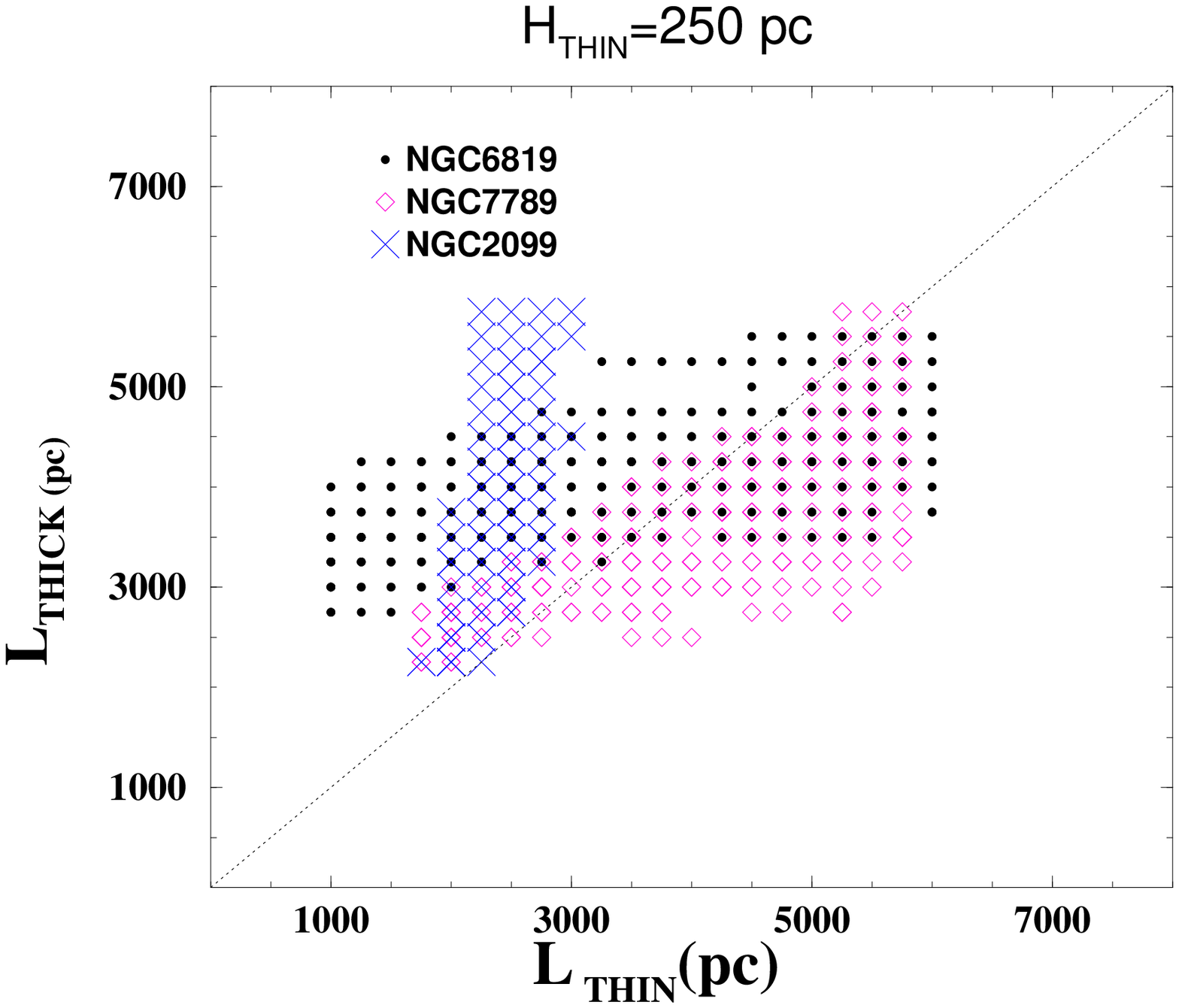}
\includegraphics[width=7cm]{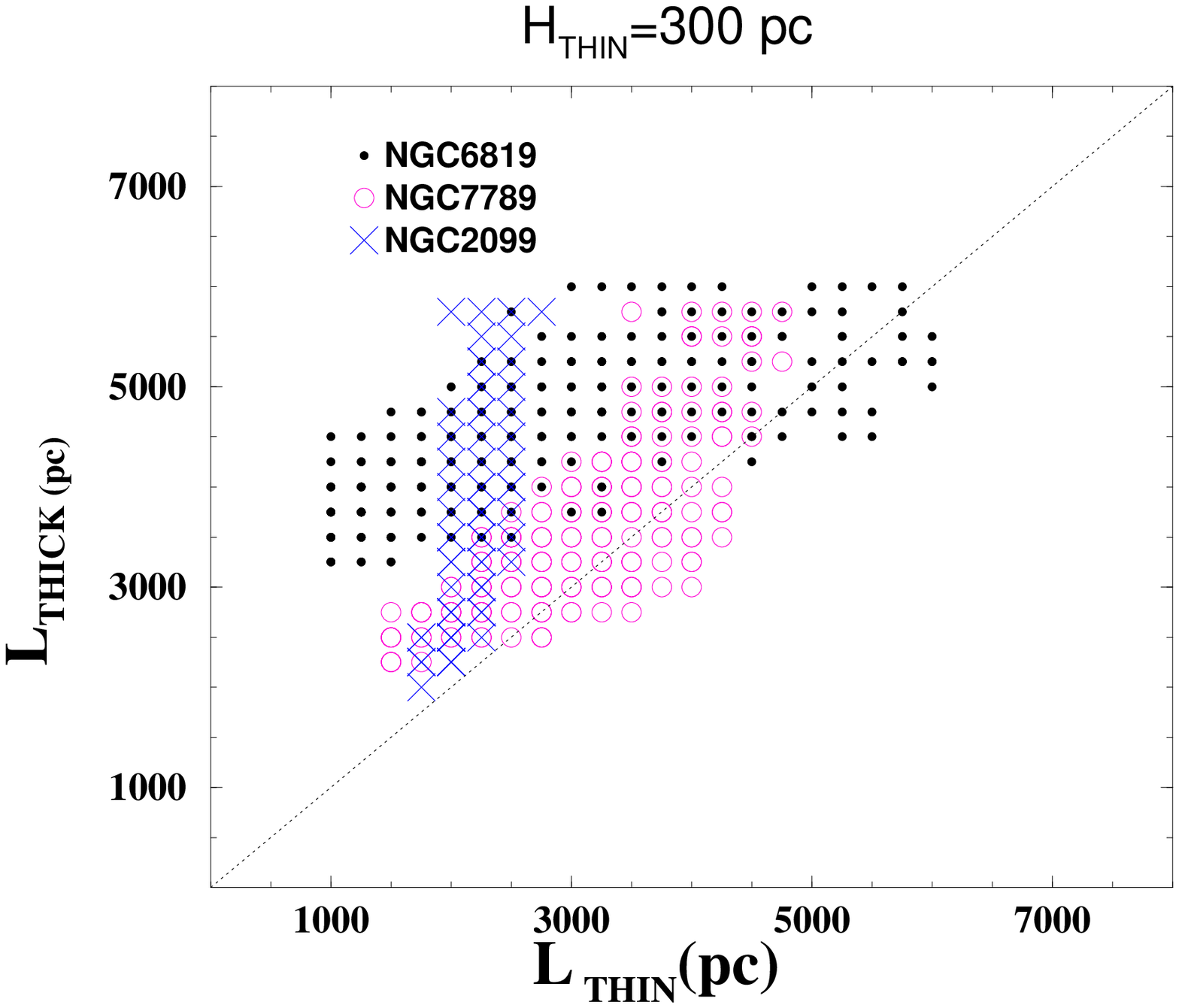}
\caption{Visual inspection of the acceptable models: different symbols
  represent different directions. From top to bottom the thin disc scale
  height is assumed as indicated.}
\label{res1} 
\end{figure}

\begin{figure}
\centering
\includegraphics[width=7cm]{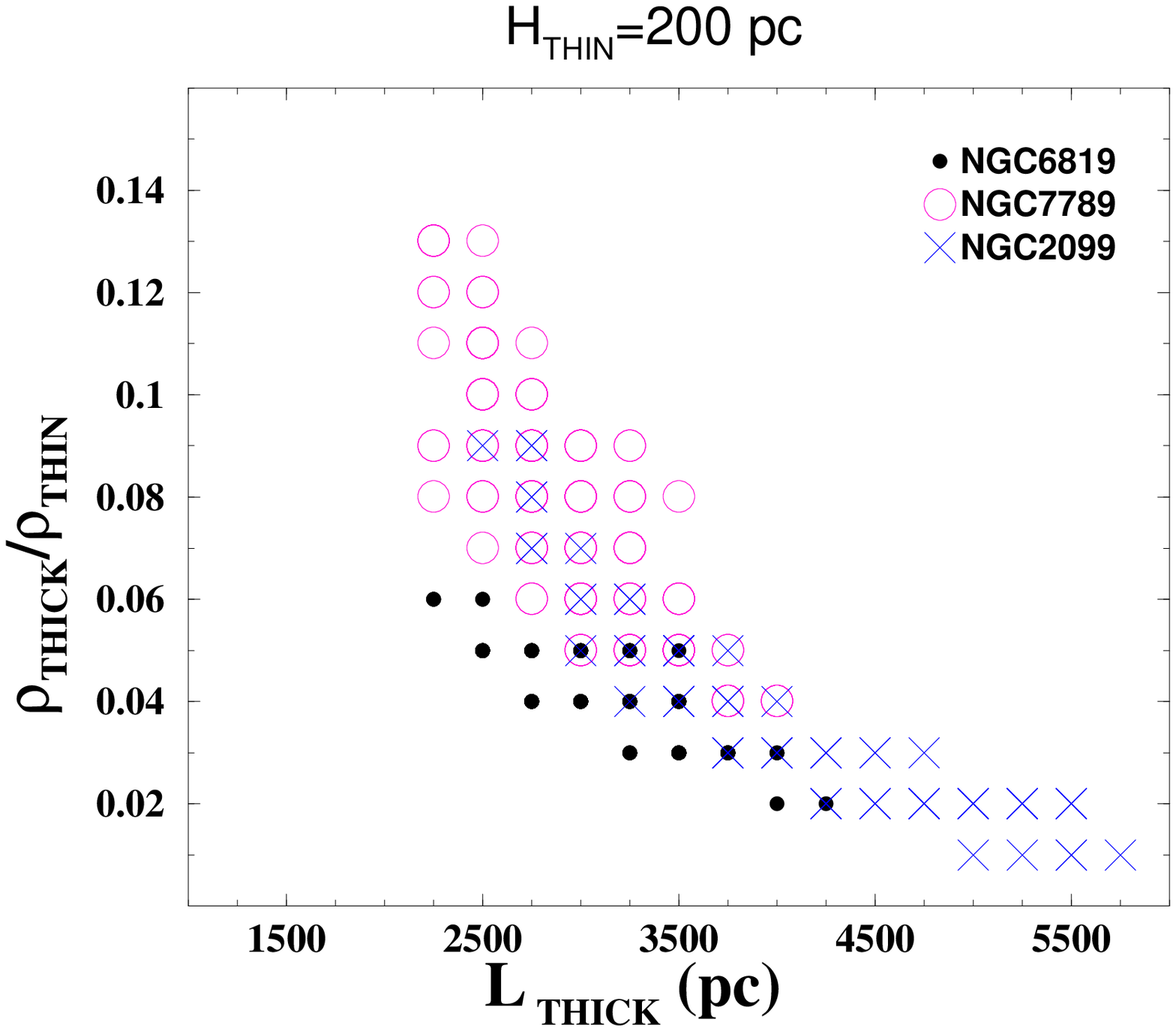}
\includegraphics[width=7cm]{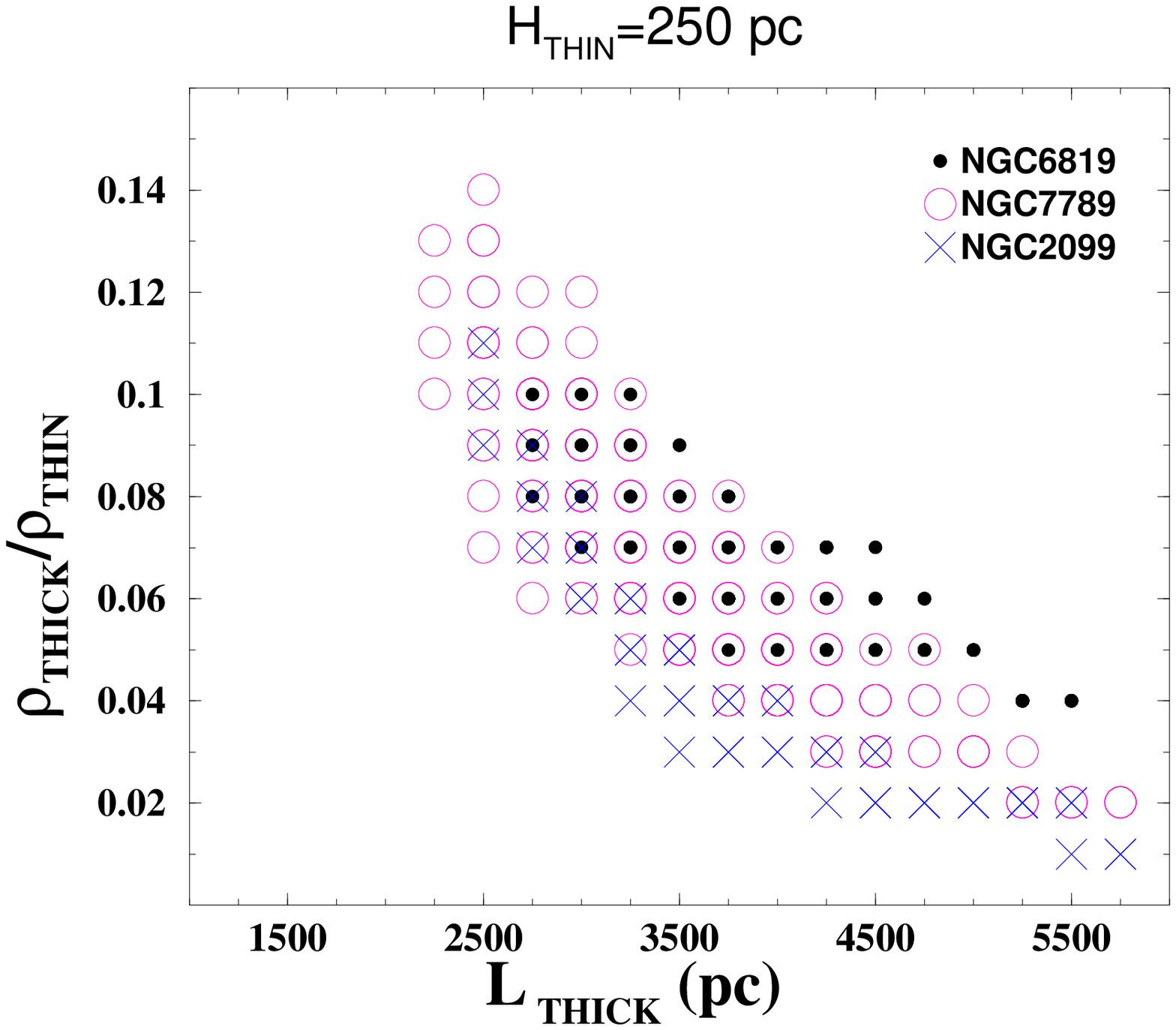}
\includegraphics[width=7cm]{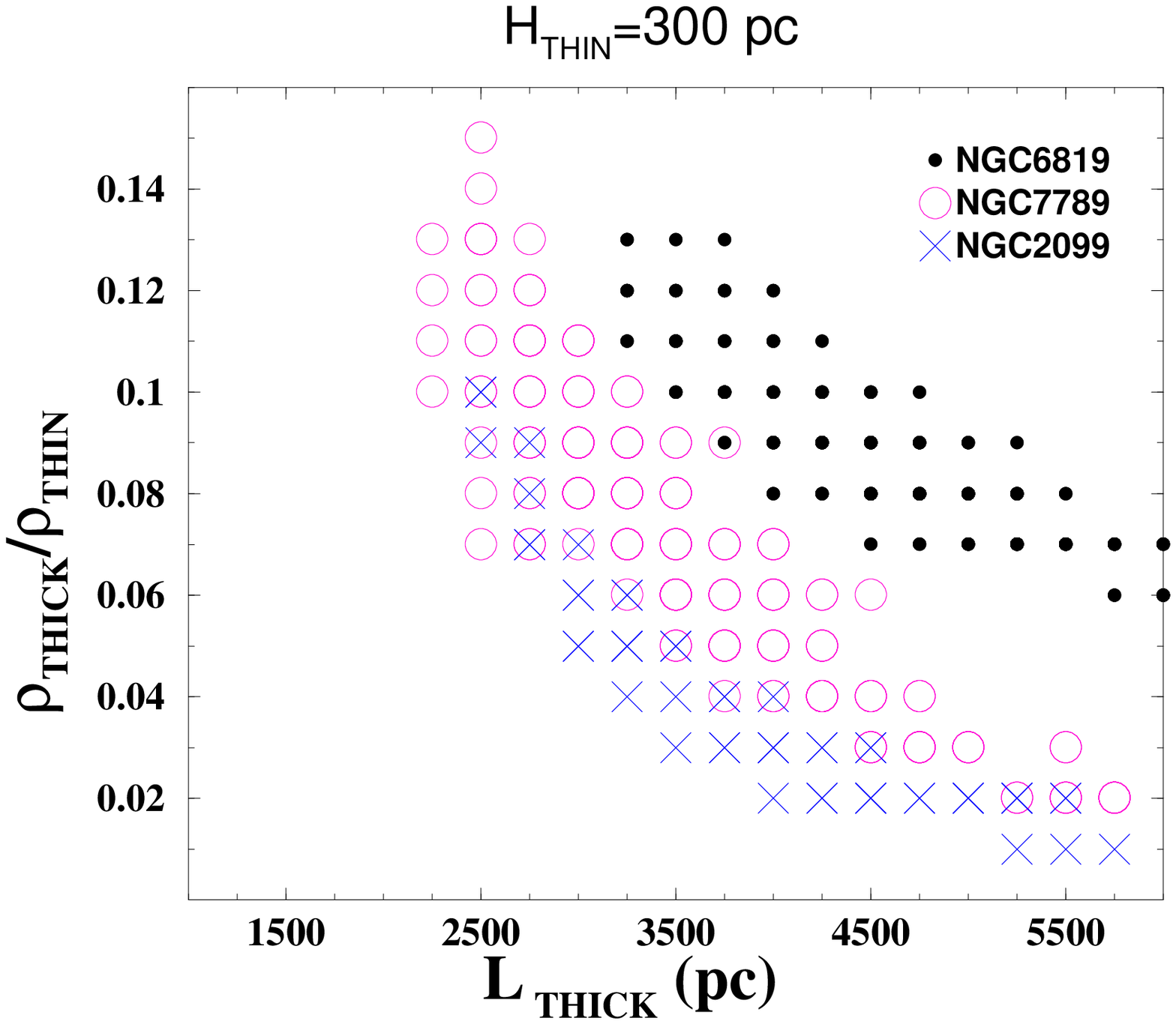}
\caption{Acceptable pairs of local thick disc normalization and thick disc
  scale length.}
\label{res2} 
\end{figure}

\section{Reddening and star formation rate}
Together with the spatial structure, the reddening distribution and the thin
 disc star formation are also constrained. For this task, particularly
 informative is the blue edge of the CMDs, namely the envelope of the main
 sequence turn-offs (vertical or shifted to the red by reddening). Once the
 model metallicity is assumed, the blue edge of the colour-magnitude diagram
 is a function of the star formation rate and the reddening distribution along
 the line of sight. In particular, the brightest stars of the blue edge (i.e.,
 the blue plume to the left and/or immediately below the cluster turn-off) are
 ideal candidates to infer information on the thin disc SFR: given the
 proximity of these stars, it is reasonable to suppose a low reddening.
 Figures \ref{fig1} and \ref{fig2} show clearly that for NGC6819 and NGC7789
 the blue edge is quite constant in colour, with $B-V$ ranging between 0.6 and
 0.7. This feature is a strong clue that in these directions the reddening is
 fairly independent of distance. For NGC2099 the situation is different,
 with the blue envelope moving from $B-V \sim 0.4$ to $B-V \sim 1.15$, so the
 reddening is expected to vary along the line of sight. In order to infer
 the thin disc SFR and the reddening distribution, the synthetic diagrams are
 computed from the following principles:

\begin{enumerate}
\item{ Given that our data are only weakly constraining the precise star
formation law, the SFR is assumed constant and \emph{only the recent SFR
cut-off is allowed to vary};}
\item{The models have been reddened using the standard $R_V =3.1$ reddening
curve of \cite{card}. The distribution of the reddening material is
assumed to be a free function of the heliocentric distance. The reddening is
increased (with steps of 1 kpc) until the synthetic and the observed blue
edge match (according to the KS test for the colour distributions) at any
distance.}
\end{enumerate}

In general, for a young population like the thin disc, the
degeneracy between SFR and reddening is high. In contrast, for an old
population like the thick disc (age $>7-8$ Gyr) the SFR has a minor
impact\footnote{7-8 Gyr correspond to turn-off masses below $1\,M_{\odot}$,
whose evolutionary times are very long.} and the blue edge is a strong
indicator of the reddening.

In the following we report our results for each field.

\subsection{$(73.98,\,\,+8.48)^\circ$ ({\textit NGC6819})}

In order to reconcile the theoretical thick disc turn-off with the observed
blue edge, our analysis yields a constant reddening $E(B-V)\sim 0.16$, which is roughly
the same as that of the cluster (according to \citealt{kato}, E(B-V) is
$0.1-0.15$ and the distance from the Sun $\sim 2.5 $ kpc), whereas it is lower
than Schlegel's (1998) estimate, E(B-V)=0.2. This result is a consequence
of the relatively ``high'' Galactic latitude (b=8.5), which confines the
dimming material very close to the Sun\footnote{At 3.0 kpc the line of sight
is higher than 450 pc over the Galactic plane, thus, exceeding the vertical
structure of the disc, it is plausible that most of the extinction occurs
before the cluster.}.

For a similar reason, we also find that a thin disc star formation which is
still ongoing is apparently inconsistent with the observed CMD: indeed, to
reproduce the blue edge at magnitudes fainter than V=18, the thin disc
activity must be switched off about 1-1.5 Gyr ago: evidently, most of the
sampled thin disc stars belong to the old thin disc, while younger objects
(age $<1$ Gyr), with a typical scale height of 100 pc, are severely
under-sampled\footnote{The solid angle and the Galactic latitude conspire to
strongly reduce the number of visible stars from the young component.} in our
field. To further examine and verify this hypothesis, we have included in our
Galactic model a synthetic young population with an exponential scale height
of 100 pc: although the presence of the cluster in the CMD does not allow a quantitative
comparison, we have verified that this intrusion has a minimum impact in the
range $18<V<22$. In other words, ongoing star formation very close to the
Galactic plane is not actually ruled out.

\subsection{$(115.5,\,\,-5.38)^\circ$ ({\textit NGC7789})}
As for NGC6819, a constant reddening correction, namely E(B-V)=0.2 for all the
synthetic stars (i.e., independently of the distance), gives the best
result. This value, while in good agreement with the cluster
estimate\footnote{According to the WEBDA database, reddening and distance for the
cluster NGC7789 are 2300 pc and E(B-V)=0.22 respectively.}, is signicantly
lower than that quoted on Schlegel's maps [E(B-V)=0.4] in this direction. As
with NGC6819, thin disc star formation seems extinguished about 1-1.5 Gyr ago,
but again this is a consequence of the selection effect against young,
low-latitude stars.

\subsection{$(177.63,\,\,+3.09)^\circ$ ({\textit NGC2099})} This field lies in the Galactic plane, and therefore we expect
the reddening to be a function of the heliocentric distance. This is indeed
clear from the CMD of Figure \ref{fig3} (upper panel); while CMDs of the other
fields show a quasi vertical blue edge, as a consequence of the reddening
material along the line of sight, the blue edge in this field is redder at
fainter magnitudes. According to simulations, this field is thin disc
dominated. A constant SFR which is still ongoing is necessary to explain the
closest stars. If the same SFR is assumed for the entire thin disc, the
reddening law we recover is indicated in Figure \ref{2099red} (the error
bars indicate the range of acceptable models). For comparison, the same figure
shows also the theoretical reddening distribution, for the same direction,
expected using an exponentially decaying law perpendicular to the Galactic
plane (the asymptotic value is fixed to the recovered one). The recovered
distribution is not fitted by any exponential reddening (calculations for
$H_{RED}=$ 100 and 200 pc are showed).

\begin{figure}
\centering
\includegraphics[width=8cm]{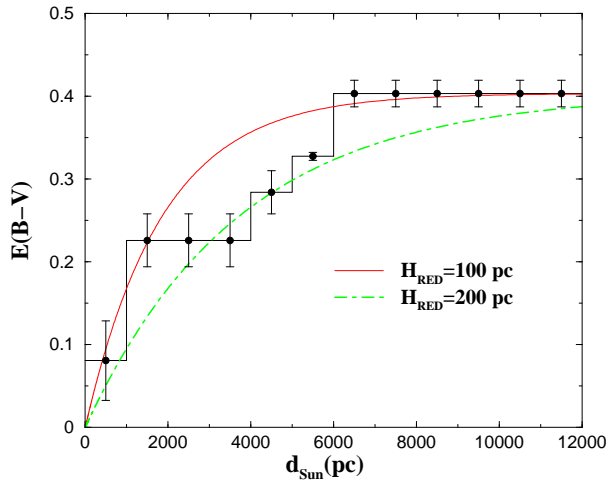}
\caption{Recovered $E(B-V)$ vs heliocentric distance for the direction of
    NGC2099 [(\textit{l,\,b})$^\circ$ = $(177.63,\,\,+3.09)^\circ$]. For
    reference, exponential reddening distributions are also plotted (the
    scales are 100 and 200 pc respectively).}
\label{2099red} 
\end{figure}

\section{Remaining uncertainties}

Uncertainties in Galaxy modelling due, for example, to the assumed
metallicity, binary population and incompleteness, can make our estimates
imprecise:

\emph{Metallicity ---} The mean metallicity for both the thin and the thick disc
is debated. Recent and old observational studies on the solar neighbourhood
provide evidence for a metallicity spread (see e.g. \citealt{nord}) at any given age which is not explained in the framework of standard models of chemical
evolution. Moreover, the presence of an age-metallicity relation and/or a
spatial metallicity gradient further complicate the issue. Ignoring the right
metallicity can lead to an incorrect interpretation of the spatial
structure. Following the paradigm that a metal rich system appears
under-luminous with respect to a metal poor one, if the data are metal poor
compared to the model, the inferred scale length will be too short.

\emph{Binaries ---} The inclusion of binaries increases the luminosity of the
population. Our model adopt only single stars, thus, if the binary fraction is
conspicuous, the retrieved scale will turn out to be too short.

\emph{Completeness ---} Losing faint stars leads to spuriously short scale
lengths (especially for the thick disc).

\emph{Halo component ---} Standard Galactic halo parameters have been
used. However, the literature also documents very different prescriptions. A
different choice could affect the thick disc distribution.

\emph{Structure ---} Our findings make sense within the context of our
parametrisation: exponential profiles, whose radial and vertical scales are
constant within the Galaxy. Introducing new features, such as an increase of
the scale height (flare) or a warp of the Galactic plane, may produce a very
different picture. Nevertheless, our parametrisation is sufficient to
reproduce the observations. In addition, our Galactic model is
axisymmetrical. Therefore, any variation with Galactic longitude is
missed. Finally, the thick disc scale height is assumed to play a minor role
(because of the low-latitude), hence it is fixed at a canonical 1
kpc. However, sporadic strong variations from this value are reported in the
literature. In order to explore also this parameter, additional and preferably
higher latitude fields would be needed.

\section{Conclusions}

We have carried out preliminary modelling of the observations in three deep
 and low latitude Galactic fields imaged at the CFH telescope. Using a
 population synthesis method we gain information about the spatial structure,
 the thin disc star formation and the reddening distribution along the lines
 of sight. The directions (\textit {l,\,b})$^\circ$ =
 $(73.98,\,\,+8.48)^\circ$ ({\textit NGC6819}) and (\textit {l,\,b})$^\circ$ =
 $(115.5,\,\,-5.38)^\circ$ ({\textit NGC7789}) are very sensitive to the thick
 disc scale length, whereas the line of sight (\textit {l,\,b})$^\circ$ =
 $(177.63,\,\,+3.09)^\circ$ ({\textit NGC2099}) is sensitive to the thin disc
 scale length. We decompose the Galaxy into halo, thick disc, and thin
 disc. Solutions common to all lines of sight do exist and require that the
 thin disc has a vertical scale shorter than about 250 pc. The inferred radial
 scales are consistent with the thick disc equally extended or slightly larger
 than the thin disc. Our results support a typical scale of 2250-3000 pc for
 the thin disc and 2500-3250 pc for the thick disc. Similar scales for the
 thin disc are basically found by \cite{rup} ($L_{thin}\sim 2.3\pm 0.1$ kpc)
 and \cite{robin92} ($L_{thin}\sim 2.5$ kpc). It is noteworthy that the
 \cite{robin92} field is in the same direction of NGC2099, giving an
 independent support to our result (they use a different statistical
 procedure). However, these authors argue for a thin disc cut-off at 15 kpc,
 whereas our determination invokes a reddening effect. In fact, the only
 acceptable solutions for the direction to NGC2099 {\emph requires} a
 structured reddening distribution, which seems to differ from simple
 exponential laws, revealing the inadequacy of classical distributions close
 to the Galactic plane.

Concerning the thick disc, our recovered scale length is perfectly compatible
with \cite{robin} ($L_{thick}\sim 2.8\pm 0.8$ kpc), \cite{buser}
($L_{thick}\sim 3\pm 1.5$ kpc), \cite{cabrera} ($L_{thick}\sim 3.04\pm 0.11$
kpc). \cite{ojha} finds a thick disc scale length ($L_{thick}\sim
3.7\pm^{0.8}_{0.5}$ kpc) longer than ours, but the mean
metallicity they adopt is higher ($[Fe/H]\sim -0.7$). Likewise, the very
extended thick disc ($L_{thick}>4$ kpc) quoted by \cite{larsen}, may
be due to adopting the luminosity function of 47 Tucanae ($[Fe/H]\sim
-0.7/-0.8$), and could be reconcilable with our findings as well. Moreover,
our conclusion about the thick disc normalization in the solar vicinity is
consistent with the results obtained by the above-mentioned papers.

Searching for common solutions, the results presented here depict a thick disc
scale length that may be only slightly longer ($\sim 20\%$) than the thin disc
one. If the two discs are really decoupled, the task for the future will be to
understand the underlying mechanisms which have promoted very different scale
heights, while preserving similar scale lengths.

Clearly this is only a pilot study showing what could be achieved by combining
deep high quality photometric fields with appropriate models for the Galaxy's 
components. To reach a better understanding of the Galactic structure, more
fields of this kind should be studied, possibly including symmetrical
locations.
\section*{Acknowledgments}
We warmly thank Michele Bellazzini for many interesting and clarifying
discussions and the referee Blair Conn for the encouraging and constructive
comments. This work is partly funded through PRIN-INAF-2005.  Based on
observations obtained at the Canada-France-Hawaii Telescope (CFHT) which is
operated by the National Research Council of Canada, the Institut National des
Sciences de l'Univers of the Centre National de la Recherche Scientifique of
France, and the University of Hawaii.

\end{document}